\documentstyle[preprint,aps,epsf]{revtex}
\tightenlines
\begin{document}
\draft

\preprint{BNL-HET-98/33}
\vspace{0.2in}

\title{J/$\psi$ Suppression in Nucleus-Nucleus Collisions}

\author{Jianwei Qiu$^{1,2}$, James P. Vary$^1$ and Xiaofei Zhang$^1$}
\address{$^1$Department of Physics and Astronomy,
             Iowa State University \\
             Ames, Iowa 50011, USA \\
         $^2$Physics Department, Brookhaven National Laboratory\\
             Upton, New York 11973-5000, USA }

\date{September 17, 1998}

\maketitle

\begin{abstract}
We propose a model for the production and suppression of J/$\psi$
mesons in high-energy hadronic collisions.  We factorized the
process into a production of the $c\bar{c}$ pairs of relative
momentum $k=k_c-k_{\bar{c}}$ convoluted with a transition probability
from the produced $c\bar{c}$ pairs into the observed J/$\psi$ mesons.
As the produced $c\bar{c}$ pairs exit the nuclear matter,
the multiple scattering between the colored $c$ and $\bar{c}$ and the
nuclear medium increases the square of the relative momentum between
the $c$ and $\bar{c}$,
$q^2=-k^2$, such that some of the $c\bar{c}$ pairs gain enough
invariant mass to transmute into open charm states.  With
only one parameter, the amount of energy gained by the produced
$c\bar{c}$ pair per unit length in the nuclear medium, our
model can fit all observed J/$\psi$ suppression data including recent
NA50 data from Pb-Pb collisions. 
\end{abstract}
\vspace{0.2in}



The suppression of J/$\psi$ production in high energy nucleus-nucleus
collisions has been suggested as a potential signal for the existence
of the quark-gluon plasma \cite{MS}.  In recent years, strong J/$\psi$
suppression has been observed in high energy hadron-nucleus and
nucleus-nucleus collisions by a number of experiments
\cite{NA3,E772,NA38}; and various theoretical explanations of the
observed J/$\psi$ suppression have been proposed \cite{Dima}.
Recently, the NA50 Collaboration at CERN observed a much stronger
J/$\psi$ suppression in Pb-Pb collisions at SPS energies \cite{NA50}.
It has been argued that a ``conventional'' approach cannot explain
this new data, and debates and controversies on the theorized origin
of the suppression have reached a new level \cite{Dima}.  In this
letter, using the same mechanism proposed in
Ref.~\cite{BQV} for the J/$\psi$ suppression in
hadron-nucleus collision, we explain the newly observed strong
J/$\psi$ suppression in high energy Pb-Pb collisions.

The production of J/$\psi$ mesons in high energy hadronic collisions
is believed to have two factorizable stages: the production of
$c\bar{c}$ pairs, and the formation of J/$\psi$ mesons from the
produced $c\bar{c}$ pairs.  Due to the large mass of
the charm quark, the $c\bar{c}$ pair production should be a
short-distance process in nature, and therefore calculable
with perturbative methods.  On the other hand, the formation of
J/$\psi$ mesons from the initially compact $c\bar{c}$ pairs
takes a relatively long time and is nonperturbative \cite{BM}.
Throughout the years, the main debate on the mechanism of J/$\psi$
production has focused on the second stage.
Three models are commonly used in the literature
for calculating the cross sections of J/$\psi$ production: the Color
Evaporation Model (CEM) \cite{Evap}, the Color-Singlet Model (CSM)
\cite{Singlet}, and the Color-Octet Model (COM) \cite{Octet}.
Since the data on J/$\psi$ suppression in nucleus-nucleus collisions
are presented in terms of total cross sections \cite{NA38,NA50},
we will focus the remainder of our discussions on the total production
rate of J/$\psi$ mesons in hadronic collisions at fixed target
energies.  For calculating the \underline{total} 
J/$\psi$ production rate,
we propose a simple model which is general enough to admit any of the
three possible formation mechanisms mentioned above.

For the collisions between hadrons (or nuclei) $A$ and $B$,
$A(p_A) + B(p_B) \rightarrow \mbox{J/}\psi(P_{{\rm J/}\psi}) + X$,
the total J/$\psi$ cross section can be factorized as follows,
\begin{eqnarray}
&&\sigma_{A+B\rightarrow {\rm J/}\psi +X}(S)
= \int dq^2 \int d^4Q \left(
  \frac{d\sigma_{A+B\rightarrow c\bar{c}+X}}{d^4Q}\right)
\nonumber \\
&& \quad\quad\quad\quad
\times
  \delta\left(q^2+(k_c-k_{\bar{c}})^2\right)
  F_{c\bar{c}\rightarrow{\rm J/}\psi}(q^2) \ ,
\label{fact}
\end{eqnarray}
where the four-vector $Q^\mu=k_c^\mu + k_{\bar{c}}^\mu$ is the total
momentum of the produced $c\bar{c}$ pair.  In Eq.~(\ref{fact}),
the variable $q^2$ is equal to the square of the relative momentum
between the $c$ and $\bar{c}$ in their rest frame,
$q^2=(2\vec{k}_c)^2$.  If the $c$ and $\bar{c}$ can be approximated
as on their mass-shell, $k_c^2 = k_{\bar{c}}^2= m_c^2$, and we have
$q^2 = Q^2 - 4\, m_c^2$.  Without trying to separate contributions
from different color channels, we define the
$F_{c\bar{c}\rightarrow{\rm J/}\psi}(q^2)$ in Eq.~(\ref{fact}) to be
a transition probability for a color-averaged $c\bar{c}$ pair of
the relative momentum square $q^2$ to evolve into a physical J/$\psi$
meson.  We propose three alternatives for parameterizing the transition
probability,
\begin{mathletters}
\label{model}
\begin{eqnarray}
F^{\rm (C)}_{c\bar{c}\rightarrow{\rm J/}\psi}(q^2)
&=& N_{{\rm J/}\psi}\, \theta(q^2)\, \theta(4m'^2-4m_c^2-q^2) \ ,
\label{Fevap}
\\
F^{\rm (G)}_{c\bar{c}\rightarrow{\rm J/}\psi}(q^2)
&=& N_{{\rm J/}\psi}\, \theta(q^2)\,
\exp\left[-q^2/(2\alpha^2_F)\right] \ ,
\label{gauss}
\\
F^{\rm (P)}_{c\bar{c}\rightarrow{\rm J/}\psi}(q^2)
&=& N_{{\rm J/}\psi}\, \theta(q^2)\, \theta(4m'^2-4m_c^2-q^2)
\nonumber \\
&\times & \left(1-q^2/(4m'^2-4m_c^2)\right)^{\alpha_F}\ ,
\label{power}
\end{eqnarray}
\end{mathletters}
where $m'$ is the mass scale for the open charm threshold. 
In Eq.~(\ref{model}), $N_{{\rm J/}\psi}$ and $\alpha_F$ are
to be fixed by
fitting the existing total production cross section data from
hadron-hadron collisions.


The transition probabilities in
Eq.~(\ref{model}) represent a wide range of J/$\psi$ formation
mechanisms.  The $F^{\rm (C)}(q^2)$ implies that all $c\bar{c}$ pairs
with invariant mass below the open charm threshold have the same
\underline{constant} (C) probability to become the physical J/$\psi$
mesons, which is effectively the same as the Color Evaporation Model
\cite{Evap}.  The $F^{\rm (G)}(q^2)$ corresponds to following
assumptions: the 
transition amplitude $\langle c\bar{c} | {\rm J/}\psi\rangle$ does not
involve any radiation and interaction with the medium, and it is then
proportional to the J/$\psi$ wave function parameterized as a
\underline{Gaussian} (G).
If we neglect the $q^2$-dependence in the production of the $c\bar{c}$
pairs in Eq.~(\ref{fact}), and require the $c\bar{c}$ to be
color-singlet, the total cross section with $F^{\rm (G)}(q^2)$ is
effectively the same as that from the Color-Singlet Model
\cite{Singlet}.

If the J/$\psi$ mesons are formed after a long-time expansion from the
small size $c\bar{c}$ pairs, and radiating soft gluons adjusts the
color of the pairs, it is then natural to assume that the
$q^2$-dependence of the transition probability is associated with that
radiation, and to choose a \underline{power-law} (P)
distribution, $F^{\rm (P)}(q^2)$
in Eq.~(\ref{power}), for the transition probability. If we expand the
transition probability at $q^2\approx 0$, the normalization  of
$F^{\rm (P)}(q^2)$ can be 
related to the combination of the matrix elements in the Color-Octet
Model \cite{Octet}. 

In principle, with a different functional form of
$F(q^2)$, our factorized formula in Eq.~(\ref{fact}) can be
generalized to calculate the \underline{total} cross sections for
producing other quarkonium states.

To evaluate the J/$\psi$ total cross section in Eq.~(\ref{fact}), we
need to calculate the production rate for the $c\bar{c}$ pairs at
invariant mass $Q^2$.  As argued in Ref.~\cite{CSS}, the production
rate 
can be factorized into a convolution of two parton distributions
from the two incoming hadrons and a short-distance hard part,
$d\hat{\sigma}_{a+b\rightarrow c\bar{c}+X}/dQ^2$, which
represents the perturbatively calculable
hard parts for
the parton $a$ and $b$ to produce the $c\bar{c}$ pairs with mass $Q^2$.
Similar to the total Drell-Yan cross section, the
one-scale cross section $d\hat{\sigma}/dQ^2$ for producing the
$c\bar{c}$ pairs at the fixed target energies should be
well-represented by the leading order calculations in $\alpha_s$, and
the high order corrections are given by a smooth K-factor.  At the
leading order in $\alpha_s$, the partonic contributions 
come from two subprocesses: 
$q\bar{q} \rightarrow c\bar{c}$ and $gg\rightarrow c\bar{c}$.  
With the K-factor for
effective high order contributions, the total J/$\psi$ cross section
in Eq.~(\ref{fact}) can be written as \cite{BQV}
\begin{eqnarray}
&&\sigma_{A+B\rightarrow {\rm J/}\psi +X}(S)
=K_{{\rm J/}\psi}\,
   \sum_{a,b} \int dq^2\, \left(
   \frac{\hat{\sigma}_{ab\rightarrow c\bar{c}}(Q^2)}{Q^2}
   \right)
\nonumber \\
&&\quad\quad \times
\int dx_F\,
   \phi_{a/A}(x_a,\mu^2)\, \phi_{b/B}(x_b,\mu^2)\,
   \frac{x_a\, x_b}{x_a + x_b}
\nonumber \\
&&\quad\quad \times \
F_{c\bar{c}\rightarrow{\rm J/}\psi}(q^2) \ ,
\label{xsec}
\end{eqnarray}
where $\sum_{a,b}$ runs over all parton flavors,
and $Q^2= q^2+4m_c^2$.  Because of the two-parton final-state at the
leading order, the incoming parton momentum fractions are fixed by the
kinematics, and given by 
$x_a =(\sqrt{x_F^2 + 4Q^2/S} + x_F)/2$ and
$x_b =(\sqrt{x_F^2 + 4Q^2/S} - x_F)/2$, respectively.
In Eq.~(\ref{xsec}), the short-distance hard parts for
producing the $c\bar{c}$ pairs,
$\hat{\sigma}_{q\bar{q}\rightarrow c\bar{c}}(Q^2)$ and
$\hat{\sigma}_{gg\rightarrow c\bar{c}}(Q^2)$,
are given in Refs.~\cite{BQV,QVZ}.  In Eq.~(\ref{xsec}), the
integration limits of $x_F$ are chosen to be consistent with the data,
and the limits of $q^2$ are specified by the functional form of
$F_{c\bar{c}\rightarrow{\rm J/}\psi}(q^2)$ in Eq.~(\ref{model}).
Eq.~(\ref{xsec}) combining the transition probability defined in
Eq.~(\ref{model}) is our model for calculating the total J/$\psi$
hadronic cross sections.

\begin{figure}
\epsfysize= 7.0in
\centerline{\epsfbox{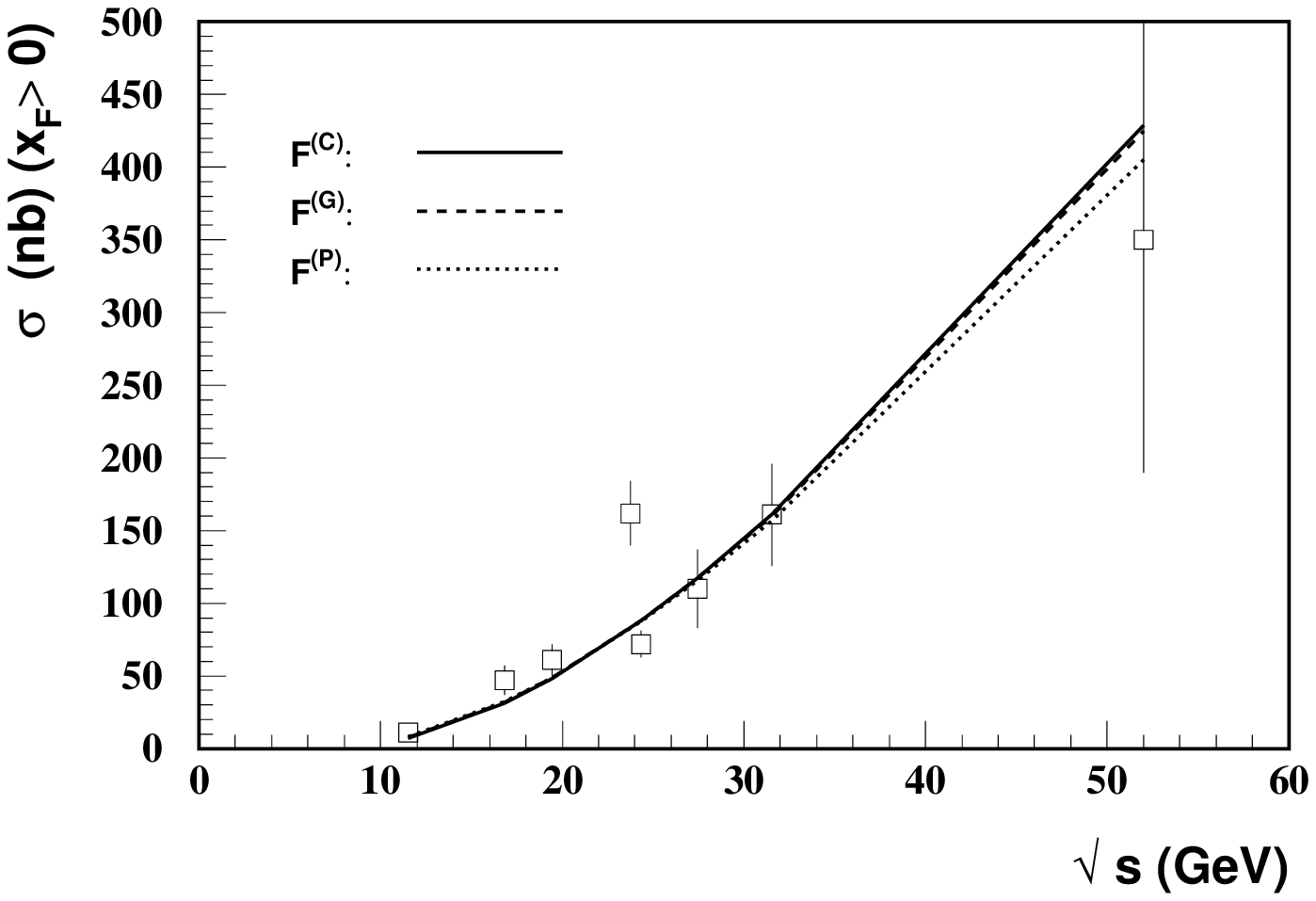}}
\caption{
Total J/$\psi$ cross sections in proton-nucleon collisions
as a function of the colliding energy
$\protect\sqrt{S}$.  }
\label{fig1}
\end{figure}

In Fig.~\ref{fig1}, we plot the total J/$\psi$ cross sections using
Eq.~(\ref{xsec}) in comparison with the data in hadronic collisions
\cite{hardP}.  Following the same fitting approach used in
Ref.~\cite{hardP}, we fix all parameters in Eq.~(\ref{model}) and list
them in Table~\ref{table1}, in which
$f_{{\rm J/}\psi} = K_{{\rm J/}\psi}\, N_{{\rm J/}\psi}$ is defined as
an overall normalization factor.  To obtain the theory curves in
Fig.~\ref{fig1}, we used CTEQ4L parton distributions \cite{CTEQ4},
and noticed that EMC effect gives a very small modification to 
the total cross
sections because of the integration of $x_F$ and $q^2$ \cite{BQV}.  In
addition, we set $m_c=1.50$~GeV and $m'=1.869$~GeV.  Choosing
different values for the $m_c$ and $m'$ changes the fitting parameters
in Table~\ref{table1} slightly.  But, it does not change the quality
of the comparison in Fig.~\ref{fig1}.
All three parameterizations in Eq.~(\ref{model})
provide a good fit to the total J/$\psi$ cross sections from
proton-nucleon collisions at fixed target energies.

In nucleus-nucleus collisions, the produced $c\bar{c}$ pairs are
likely to interact with the nuclear medium before they exit.
Observed anomalous nuclear enhancement of the momentum
imbalance in dijet production tells us that a colored parton
(quark or gluon) experiences multiple scatterings when it passes
through the
nuclear medium, and the square of the relative transverse momentum
between
two-jets increases in proportion to the size of the nucleus
\cite{Naples,LQS}.  If we let the $c$ and $\bar{c}$ be the
parent-quarks of two jets, the $q^2$ becomes the square of the
relative
momentum between the two jets in their c.m. frame.  Therefore, as
the $c$ and $\bar{c}$ pass through nuclear matter, just like a
di-jet system, the square of the relative momentum $q^2$ increases.
As a result, some of the $c\bar{c}$ pairs might
gain enough relative momentum square $q^2$ to be pushed over the
threshold to become open charm mesons, and consequently, the cross
sections for J/$\psi$ production are reduced in comparison with
nucleon-nucleon collisions. 

If the formation length for the J/$\psi$ meson, which depends on the
momenta of the $c\bar{c}$ pairs produced in the hard collision, is
longer than the size of the nuclear medium, it is reasonable to assume
that the transition
probability $F_{c\bar{c}\rightarrow {\rm J/}\psi}(q^2)$, defined in
Eq.~(\ref{model}), can be factorized from the multiple scattering.
Then, as far as the total cross section is concerned, the net effect
of the multiple scattering of the $c\bar{c}$ pairs
can be represented by a shift of $q^2$ in the transition
probability, 
\begin{equation}
q^2  \longrightarrow  \bar{q}^2 = q^2 + \varepsilon^2\, L(A,B) \ .
\label{q2shift}
\end{equation}
In Eq.~(\ref{q2shift}), $L(A,B)$ is the effective length of nuclear
medium for the $c\bar{c}$ pair to pass through in the collisions of
two nuclei of $A$ and $B$, and it depends on the
details of the nuclear density distributions \cite{LABref}. 
In Eq.~(\ref{q2shift}), the $\varepsilon^2$ represents the square of
the relative momentum received by the $c\bar{c}$ pairs per unit
length of the nuclear medium.  The value of the $\varepsilon^2$ can be
estimated from the observed nuclear enhancement in the momentum
imbalance of two-jets in hadron-nucleus collisions \cite{QVZ}.
Using the data from pion-nucleus collisions \cite{Naples},
we estimate that $\varepsilon^2 \sim 0.2-0.5$~GeV$^2$ per unit length
of nuclear medium \cite{QVZ,LQS}.

\begin{figure}
\epsfysize=7.0in
\centerline{\epsfbox{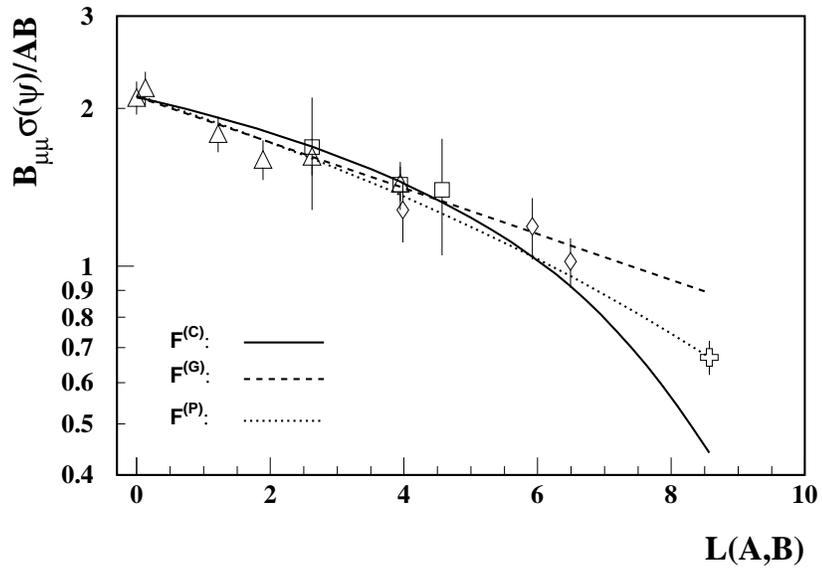}}
\caption{
Total J/$\psi$ cross sections with the branching ratio to $\mu^+\mu^-$
in proton-nucleon, hadron-nucleus and nucleus-nucleus collisions as a
function of the effective nuclear medium length $L(A,B)$. 
}
\label{fig2}
\end{figure}

In Fig.~\ref{fig2}, we plot the predictions of
J/$\psi$ total cross sections in proton-nucleon, proton-nucleus and
nucleus-nucleus collisions.  The data in Fig.~\ref{fig2} are from
Ref.~\cite{NA50_2}, in which all data were rescaled to $P_{\rm
beam}=200$~GeV.  The effective length $L(A,B)$ were chosen to be the
same as those used in Ref.~\cite{NA50_2}.  Three theory curves
correspond to three parameterizations defined in Eq.~(\ref{model}). 
The values of the parameter $\varepsilon^2$ for different
parameterizations are listed in Table~\ref{table1}, which are
consistent with our earlier estimates from the momentum imbalance of a
di-jet system.  As in Fig.~\ref{fig1}, we used the CTEQ4L
parton distributions and set $m_c=1.50$~GeV and $m'=1.869$~GeV for
plotting Fig.~\ref{fig2}.  Choosing different values for the
$m_c$ and $m'$ changes the values of $\varepsilon^2$
slightly, but, it does not change the features of Figs.~\ref{fig1}
and \ref{fig2}.

The three parameterizations of the transition probability in
Eq.~(\ref{model}) represent the different J/$\psi$ production
mechanisms, and naturally, they predict different behavior of J/$\psi$
suppression in Fig.~\ref{fig2}.  Therefore, understanding the
suppression can also help us to distinguish the production mechanism.

For the parameterization $F^{\rm (G)}(q^2)$ in Eq.~(\ref{gauss}),
a shift of $q^2$ to $\bar{q}^2$ in Eq.~(\ref{q2shift}) for the
J/$\psi$ suppression in nucleus-nucleus collisions gives following
relation 
\begin{equation}
\sigma_{AB\rightarrow {\rm J/}\psi}(S) =
\exp\left[-\frac{\varepsilon^2}{2\alpha_F^2}\, L(A,B)\right]
\sigma_{NN\rightarrow {\rm J/}\psi}(S).
\label{ABgauss}
\end{equation}
This relation is effectively the same as that predicted by the
Glauber theory, if we let the suppression factor in the simple Glauber
theory be $\exp[-\sigma_{\rm abs}\, \rho\, L(A,B)]$, with $\rho$ being
the nuclear density.  With the parameters in Table~\ref{table1}, we
have the effective absorption cross section $\sigma_{\rm abs}\sim
5.9$~mb, which is the same as that obtained in Ref.~\cite{NA50_2}.
In addition, our model interprets the $\sigma_{\rm abs}$ in Glauber
theory is proportional to the energy absorbed by the colored
$c\bar{c}$ pairs.  As expected \cite{Dima}, like the Glauber theory,
the parameterization of $F^{\rm (G)}(q^2)$ does not generate enough
suppression for heavy nucleus-nucleus collisions.  As discussed
earlier, the Gaussian parameterization corresponds to assuming that
the formation process does not have radiation, and is then
proportional to 
the square of the J/$\psi$ wave-function.  However, as we learned from
the success of the Color-Octet Model \cite{Octet}, the formation of
J/$\psi$ should involve the expansion of
the $c\bar{c}$ pairs and the radiation of soft gluons.  Therefore, we
should prefer a power-law parameterization than a Gaussian
parameterization, and consequently, we expect to have more suppression
than that expected from the Glauber approach.

Since the $F^{\rm (C)}(q^2)$ is the same as the $F^{\rm (P)}(q^2)$
when $\alpha_F=0$, we expect a maximum suppression from the
Color-Evaporation Model, as seen in Fig.~\ref{fig2}.
The Color-Evaporation Model assumes that all $c\bar{c}$ pairs with
invariant mass less than the open charm threshold should have the same
transition probability to become the J/$\psi$ meson.  However, the
phase
space cutoff for the $c\bar{c}$ pairs to become the open charm in
Eq.~(\ref{xsec}) appears classical.  In quantum theory,
the $c\bar{c}$ pairs with invariant mass less than the $4m'^2$ should
have a small, but non-zero, probability to become open charm systems,
and in general, the pairs just below the threshold should have a
relatively larger probability for a transition to open charm than those
far below the threshold.  With these considerations, we believe that
the power-law parameterization $F^{\rm (P)}(q^2)$ with $\alpha > 0$
represents the more accurate physics for the J/$\psi$ production.

Our model of the J/$\psi$ production relies on the factorization
between the production of the $c\bar{c}$ pairs and the formation of
the J/$\psi$ mesons, and accurate calculation for the production of
the $c\bar{c}$ pairs.  For the total cross sections at fixed
target energies, we believe that such factorization is justified, and
the perturbative calculations are reliable.  However, at collider
energies, most J/$\psi$ events are measured at large transverse
momentum, $Q_T$; and the $\sqrt{S}$ as well as the $Q_T$ are much
larger than the invariant mass of the $c\bar{c}$ pairs, $Q$, in
Eq.~(\ref{xsec}).  Therefore, the perturbative calculations of
$d\hat{\sigma}/dQ^2$ and $d\hat{\sigma}/dQ^2dQ_T^2$
for the production of the $c\bar{c}$ pairs
become very nontrivial due to the large logarithms
$\log(1/x) \sim \log(S/Q^2)$ and $\log(Q_T^2/Q^2)$. 
A resummation of such large logarithms to all order in $\alpha_s$ is
necessary in order to have a reliable prediction.  We defer our
detailed discussions on J/$\psi$ production and
suppression at the collider energies to another publication.

Our predictions of the J/$\psi$ suppression, as shown in
Fig.~\ref{fig2}, depend on an additional assumption: the separation of
the multiple scattering of the $c\bar{c}$ pairs and the formation of
the J/$\psi$ mesons.  We believe that this additional assumption can
only be justified when the J/$\psi$ formation length is larger than
the effective medium length $L(A,B)$ in our Eq.~(\ref{q2shift}).
Once the J/$\psi$ meson is formed, the multiple scattering with
nuclear medium should be reduced due to the color singlet nature of
the meson, and then, the Glauber formalism for the suppression
should be more relevant.  Therefore, if there is no
QCD phase transition to the quark-gluon plasma, we expect the
following
features for the J/$\psi$ suppression in nucleus-nucleus collisions.
As the size of colliding nuclei increases, the J/$\psi$ suppression
should follow the dotted curves in Fig.~\ref{fig2}; and when the
$L(A,B)$ is compatible to the J/$\psi$ formation length, the
suppression will become smaller than what is predicted by the dotted
curve. 
Finally, we conclude that our simple model for the total cross
sections of J/$\psi$ production in nucleus-nucleus collisions, as
defined in Eq.~(\ref{xsec}), can explain the existing data in
hadron-hadron, hadron-nucleus and nucleus-nucleus collisions.
Although our model is different in many aspects from what have been
discussed in the literature, we believe that it has
the key physical mechanisms for the J/$\psi$ production and
suppression.  Detailed comparison between our work and others in the
literature are given in Ref.~\cite{QVZ}.

\section*{Acknowledgment}

One of us (Qiu) is pleased to acknowledge stimulating discussions with
S. Gardner, D. Kharzeev and A.H. Mueller.
This work was supported in part by the U.S. Department of Energy under
Grant No. DE-FG02-87ER40731 and the Contract No. DE-AC02-98CH10886. 




\begin{table}
\caption{Values of parameters used to produce the theory curves in
Figs.~\protect\ref{fig1} and ~\protect\ref{fig2}.}
\label{table1}
\begin{tabular}{c|cc|cc|cc}
                  & $F^{\rm (C)}$ & & $F^{\rm (G)}$ & & $F^{\rm (P)}$ &
\\ \hline
$f_{{\rm J/}\psi}$&  0.248        & &    0.470      & & 0.485         &
\\ \hline
 $\alpha_F$       &   0           & &   1.2 GeV     & & 1.0           &
\\ \hline
 $\varepsilon^2$ (GeV$^2$/fm) 
                  &  0.45         & &  0.29         & & 0.25          &
\end{tabular}
\end{table}

\end{document}